\documentclass[preprint,authoryear,12pt]{elsarticle}

\usepackage{amssymb}

\journal{Thin Solid Films}

\begin{document}

\begin{frontmatter}



\title{Thin films thickness Measurement by the conductivity theory in the framework of born approximation}


\author[label1]{M. Jannesar}
\author[label1]{G. R. Jafari}
\author[label2]{S. Vasheghani Farahani}
\author[label1]{S. Moradi}

\address[a]{Department of Physics, Shahid Beheshti University, G.C., Evin, Tehran 19839, Iran}
\address[b]{Centre for Plasma Astrophysics, Department of Mathematics, Katholieke Universiteit Leuven, Celestijnenlaan 200B bus 2400, B-3001 Heverlee, Belgium}

\begin{abstract}
When the thickness of the layer is smaller than the electrons mean
free path, the morphology affects the conductivity directly based on
the layer thickness. This issue provides basis in order to estimate
the thickness of the layer by understanding the morphology and the
value of the conductivity. This method is an inverse approach on
thickness estimation and is applied to various samples. The
comparison of the results with other thickness estimations shows
good consistency. The benefits of this approach is that the only
parameter that needs to be measured is the conductivity, which is
quite trivial. Despite the simplicity of this approach, its results
would prove adequate to study both the material properties and the
morphology of the layer. In addition, the possibility of repeating
the measurements on thickness for AC currents with various
frequencies enables averaging the measurements in order to obtain
the most precise results.
\end{abstract}

\end{frontmatter}


\section{Introduction}

A wide range of studies have been carried out on the conductivity
properties of thin films in the past. A very popular aspect of
studying electric conductivity of thin films is its dependence on
the surface \cite{tho}. The first classical formulation in this
context was proposed by \cite{fuch} and generalized considering the
presence of a magnetic field by \cite{sond}. The logic behind
considering the magnetic field \cite{kao} is that the thickness of
the thin film (e.g. Indium) would affect the critical magnetic
field, consequently affecting the transition from the
super-conducting state to the normal state \cite{tox}. A lot of
effort has been invested on surfaces in the presence of magnetic
fields in the context of e.g. surface quantum states and impedance
oscillations \cite{nee,kha}, electron scattering \cite{koc,che} and
transport (conductivity) \cite{gol,fi1,pa1,pal,mey}, etc, which
where all based on the quantum mechanical formulation \cite{pra}.
Quantum size effects and surface roughness influences the
conductivity of ultra-thin magnetic layers \cite{br1,br2,jal,hof}
and wave scattering \cite{wave1,wave2}.

A general expression was suggested for the dependence of the
conductivity on the thin film thickness \cite{fi1}, where the
surface roughness was characterized by the autocorrelation function
that described all the electrical properties of the system. This
enabled discussion on the interplay of the correlation length of the
surface and the Fermi wave vector with regards to the conductivity
\cite{fi2}. Note that in this model the thin films thicknesses is
smaller than the bulk electrons' mean free path. It was deduced that
the conductivity only depends on the thickness of the film provided
that $\xi k_{F}\ll1$, where $\xi$ and $k_{F}$ are the correlation
length of the surface and the Fermi wave vector, respectively. But
on the other hand, for the case $\xi K_{F}\gg1$, the autocorrelation
function would act effectively on the conductivity of the thin film
\cite{fi1,fi2}.

In a series of studies, Palasantzas extended the previous models by
pointing out that the film surface ought to be considered fractal.
In a sense that the local fractal dimensions would also affect the
characteristics of the self-affine fractal model (e.g. electric
conductivity) for one rough layer \cite{pa1} or two rough layers
\cite{pa2}. Here we take an inverse approach based on the two layer
self-affine fractal model of Palasantzas \cite{pa2}, where we
measure the electric conductivity in order to obtain details on the
thickness of the layer. In this article we implement a model to
determine the thickness of metallic films with two rough boundaries
that is based on their conductivity in the framework of Born
approximation. In section II, the theory and formulation for the
conductivity based on Born approximation in presence of two rough
boundaries is illustrated. In section III, effects of conductivity
on the thickness of the layer is discussed in addition to the by
product effects of the alternate and direct currents applied to the
layer its resistance. In section IV the summary is stated.

\section{Inverse method for measuring the thin-films thickness}

Consider a thin film of two rough surfaces with thickness $d$, which
its thickness is smaller than the mean free path of the electron.
The thin film crosses the z-axis at $d/2$ and $-d/2$, with
$z_{1}(\vec{r})=d/2+h_{1}(\vec{r})$ and
$z_{2}(\vec{r})=-d/2+h_{2}(\vec{r})$, where $h_{1}$ and $h_{2}$ are
the random roughness fluctuations of surface ${1}$ and ${2}$
respectively with $\langle h(\vec{r})\rangle=0$ \cite{pa2}. Note
that the choice of the quantum mechanical formulation in this
limitation is due to the fact that the thickness layer is smaller
than the mean free path of the electron. In this case the electrons
would scatter from the surface before being scattered by the bulk
electrons \cite{pa1,pa2}. The conductivity relation for the case
where the surface roughness $\omega$ is considered much smaller than
the film thickness $d$ $(\omega\ll d)$) is \cite{pa2}
\begin{equation}
\label{1}
\sigma=(4e^{2}/hd)\sum_{\nu=1}^{N}\sum_{\acute{\nu}=1}^{N}(E_{F}-\varepsilon_{\nu})(E_{F}-\varepsilon_{\acute{\nu}})
\left[([D^{in}(E_{F})_{\nu\acute{\nu}}]+[D^{cor}(E_{F})_{\nu\acute{\nu}}])\right]^{-1},
\end{equation}
where $N$ is the number of occupied mini-bands, $E_F$ is the Fermi
energy, $D$ is the scattering matrix which contains two terms:
incoherent term $D^{in}$, (incoherent scattering by two rough
interfaces) and cross-correlation term $D^{cor}$, (coherent
scattering by two interfaces). The elements of $D^{in}$ and
$D^{cor}$ describe intra- and inter-subband transitions, which is
expressed by \cite{pa2}
\begin{equation}
\label{2}
[D^{in}(E_{F})_{\nu\acute{\nu}}]=\sum_{b=1}^{2}[\delta_{\nu\acute{\nu}}\sum_{\mu=1}^{N}
q_{\nu}^{2} L_{b}^{\nu\mu}\int_{0}^{2\pi}\langle
|h_{b}(q_{\nu\mu})|^{2}\rangle
d\theta-q_{\nu}q_{\acute{\nu}}L_{b}^{\nu\acute{\nu}}\int_{0}^{2\pi}\langle
|h_{b}(q_{\nu\acute{\nu}})|^{2}\rangle\cos\theta\ d\theta],
\end{equation}
and
\begin{equation}
\label{3}
[D^{cor}(E_{F})_{\nu\acute{\nu}}]=2[\delta_{\nu\acute{\nu}}\sum_{\mu=1}^{N}
q_{\nu}^{2} L_{12}^{\nu\mu}\int_{0}^{2\pi}\langle
|h_{12}(q_{\nu\mu})|^{2}\rangle
d\theta-q_{\nu}q_{\acute{\nu}}L_{12}^{\nu\acute{\nu}}\int_{0}^{2\pi}\langle
|h_{12}(q_{\nu\acute{\nu}})|^{2}\rangle\cos\theta\ d\theta].
\end{equation}
In Eq. (\ref{2}), $\langle |h_{b}(q)|^{2} \rangle$ is the Fourier
transform of the auto-correlation function expressed as
$C_{b}(r)=(\langle h_{b}(r)\rangle\langle h_{b}(0)\rangle)$. In Eq.
(\ref{3}) $\langle|h_{12}(q)|^{2}\rangle$ is the Fourier transform
of the cross correlation function $C_{12}(r)=(\langle
h_{b}(r)\rangle\langle h_{\acute{b}}(r)\rangle)$ with
$b\neq\acute{b}$. The wave vectors of $\nu$ and $\acute{\nu}$
miniband edges in Eqs (\ref{2}) and (\ref{3}) are expressed as
$q_{\nu}=[(2m/\hbar^{2})(E_{F}-E_{\nu})]^{1/2}$  and
$q_{\nu\acute{\nu}}=(q_{\nu}^{2}+q_{\acute{\nu}}^{2}-2q_{\nu}q_{\acute{\nu}}\cos\theta)^{1/2}$,
with $\theta$ being the angle between $q_{\nu}$ and
$q_{\acute{\nu}}$. In the case where the potential $U$ tends to
infinity, the parameters $L_{b}^{\nu\mu}$ and $L_{12}^{\nu\mu}$ are
defined as  $(\hbar^{2}/4md^{3})^{2}(\nu^{2}\mu^{2})$ and
$-(\hbar^{2}/4md^{3})^{2}(\nu^{2}\mu^{2})$ and respectively, for
details of the model see \cite{pa2}.

Consider a fractal surface \cite{bar,thinfilm,zamani} with roughness
$\omega$, correlation length $\xi$, and roughness exponent $\alpha$.
In order to obtain the conductivity (Eq. (\ref{1})) of the system,
it is essential to have the power spectrum for each surface in
addition to the power spectrum for their cross correlation
\cite{pa1,zhao}
\begin{eqnarray}
\label{5} \langle|h(q)|^{2}\rangle=\left\{\begin{array}{ccc}
      2\pi\frac{\omega^{2}\xi^{2}}{(1+aq^{2}\xi^{2})^{1+\alpha}}& for & \xi K_{F}\gg1 \\
      cte & for & \xi k_{F}\ll1  \\
       \end{array}\right\} \nonumber\\
\mathrm{with} \nonumber\\
a=\left\{\begin{array}{ccc}
         (1/2\alpha)[1-(1+aq_{c}^{2}\xi^{2})^{-\alpha} & for & 0<\alpha\leq1 \\
         (1/2)\ln(1+aq_{c}^{2}\xi^{2}) & for & \alpha=0
       \end{array}\right\}
\end{eqnarray}
It is worth mentioning that the power spectrum defined by Eq.
(\ref{5}) only works for fractal surfaces \cite{zhao}.

The aim of this work is to estimate the thickness by the measured
conductivity making use of Eq. (\ref{1}). This is an inverse method
to previous methods where the conductivity was obtained by the
measured thickness of the layer, see \cite{pa1,pa2}. Due to the fact
that for any typical rough surface where the roughness is of the
same order as the thickness, the surface parameters (e.g. surface
roughness) would be effective on the determination of the layers
thickness. Note that distinguishing the roughness of the surface
from the thickness of the layer is not a simple task, this
complicates the correct estimation of the thickness. Hence,
experimental measurements of the conductivity would be more trivial
compared to measuring the thickness.

The inverse method illustrated in this work is based on the
following procedure;

- The AFM images of the interface between the substrate and the
layer together with the AFM images of the layer and air needs to be
produced in order to find the fluctuations of the surface $h(r)$.

- From the logarithmic plot of the height-height correlation
function $\langle |h(r+z)-h(r)|^2\rangle$ \cite{zhao} the surface
parameters for each interface is obtained.

- Having in hand the surface parameters, the power spectrum could be
obtained by Eq. (4). Hence, the scattering matrix of Eqs. (\ref{2})
and (\ref{3}) could be obtained.
 - In the final stage, comparing the
experimental measurements for the conductivity with the analytical
results obtained from Eq. (\ref{1}) the thickness is obtained.

\section{Discussion and conclusions}

\begin{figure}[thbp]
\includegraphics[width=7cm,height=7.1cm]{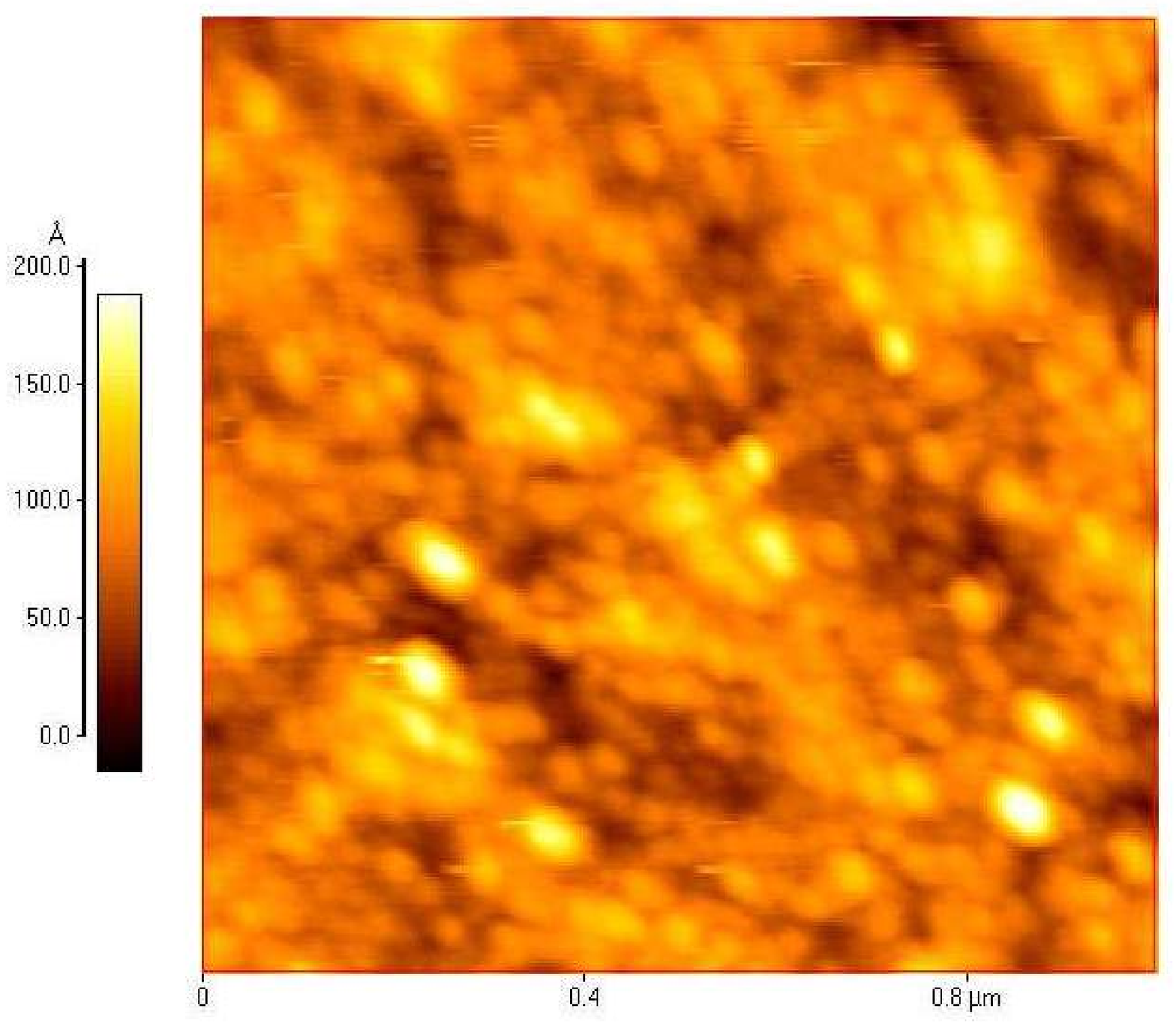}
\includegraphics[width=7cm,height=6.9cm]{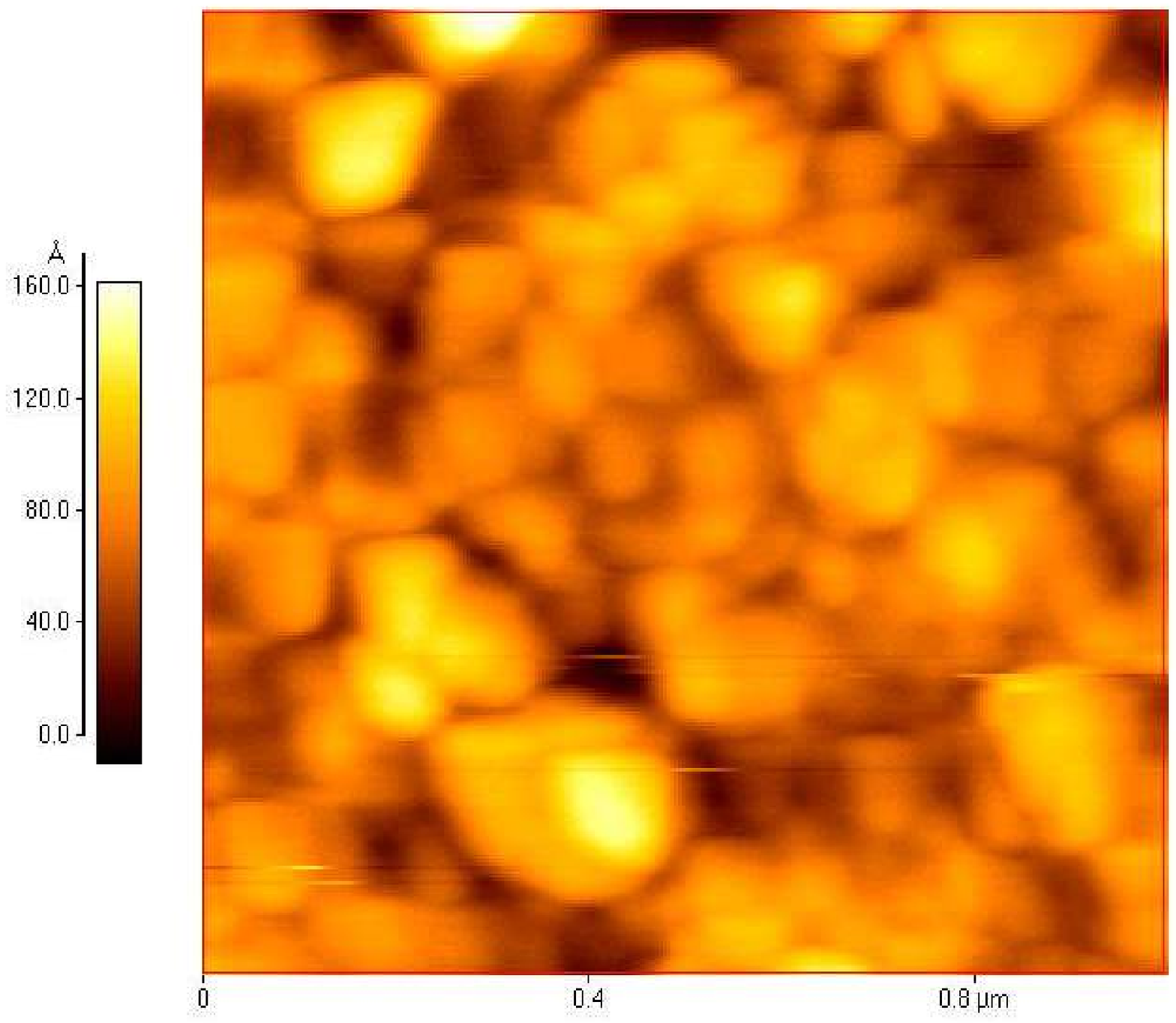}
\caption{Left panel, AFM images of the film interface with air.
Right panel, AFM images of the film interface with the substrate.}
\label{Fig1}
\end{figure}

The sample that we considered was a copper thin film that had been
coated with a coating rate of $0.9\pm0.04$, where after $8.0$
seconds, the thickness of $7.2nm$ was achieved. The topography of
the samples were investigated using Atomic force microscopy (AFM)
with Park Scientific Instruments (model Autoprobe CP). The images
were scanned into $256 \times 256$ pixels in a constant force mode
and digitised with the scanning frequency of $0.6Hz$. The AFM images
(Fig. \ref{Fig1}) are shown for both the film interface with air and
the film interface with the substrate. For this sample the thickness
of the thin film is calculated by using Eq. (\ref{1}).

\begin{figure}[thbp]
\centering
\includegraphics[width=10cm,height=8cm]{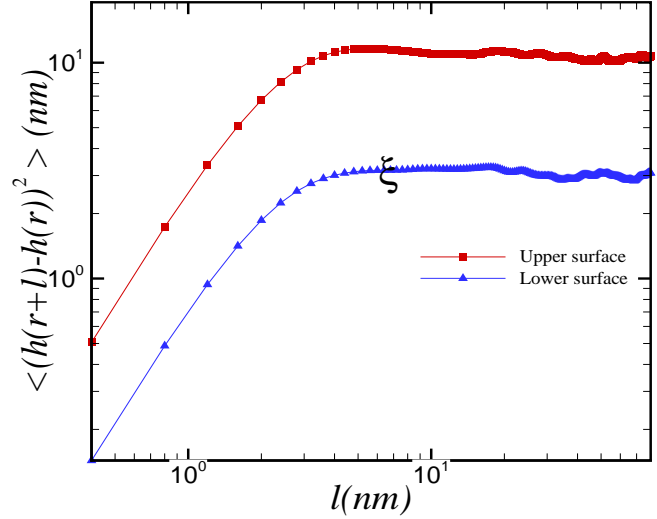}
\caption{The height-height correlation function of the interface
between the layer and substrate (bottom curve) and the interface
between the layer and air (top curve).} \label{Fig2}
\end{figure}

In order to obtain the surface parameters, the height-height
correlation function in addition to the height fluctuations of the
surface needs to be obtained. In Fig. \ref{Fig2} the logarithmic
height-height correlation function corresponding to the interface
between the layer and substrate (bottom curve) and the interfaces
between the layer and air (top curve) is plotted making use of
equation for $\langle |h(r+z)-h(r)|^2\rangle$.

The surface parameters for the copper sample under consideration is
extracted from Fig. \ref{Fig2} as
\begin{eqnarray}
\label{30} &&\xi_{b}=4 nm,\,\,\, \omega_{b}=1.1 nm,\,\,\, \alpha_{b}=0.75\nonumber \\
&&\xi_{t}=3 nm,\,\,\, \omega_{t}=2.1 nm,\,\,\ \alpha_{t}=0.8
\end{eqnarray}
The power spectrums $\langle|h_{1}(q)|^2\rangle$ and
$\langle|h_{2}(q)|^2\rangle$ which could readily be seen in Eq.
(\ref{2}) could be obtained by substituting the parameters of Eq.
(\ref{30}) in  Eq. (\ref{5}). This results in obtaining the
incoherent scattering matrix $D^{in}$. To calculate the coherent
scattering matrix (Eq. (\ref{3})) the cross power spectrum needs to
be obtained. This obliges us to plot the height-height cross
correlation function which shows to be approximately just a straight
line.

\begin{figure}[thbp]
\includegraphics[width=7cm,height=7cm]{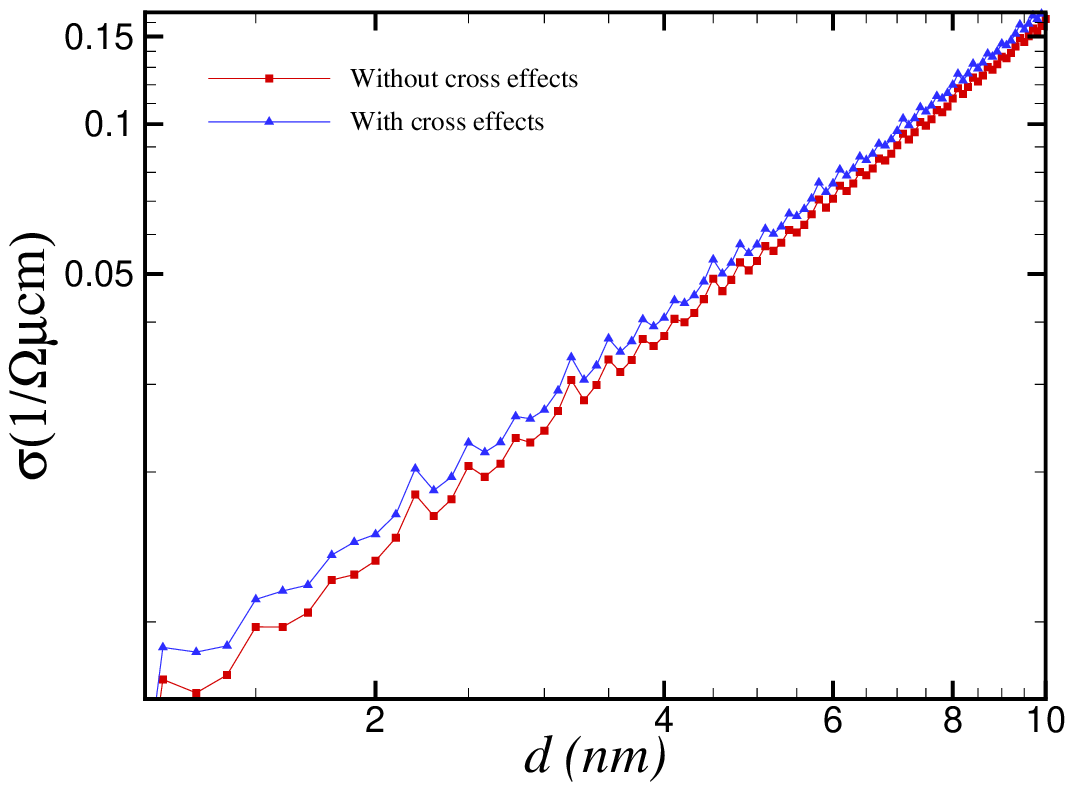}
\includegraphics[width=7.4cm,height=7cm]{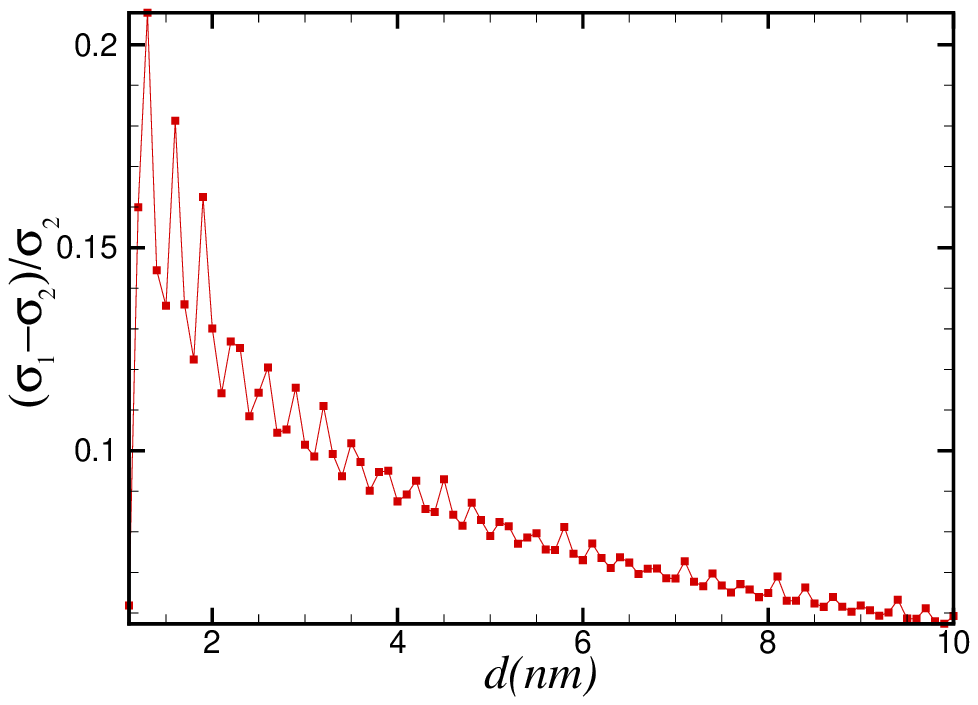}
\caption{ left panel shows the conductivity $\sigma$ vs metallic
film thickness $d$, in the presence (top curve) and absence (bottom
curve) of the cross correlation term. Right panel shows the relative
difference in conductivity as a function of thickness.} \label{Fig3}
\end{figure}

Having the values for $D^{in}$ and $D^{Cor}$, the conductivity (Eq.
(\ref{1})) could be plotted as a function of thickness, see Fig.
\ref{Fig3}. For the copper sample under study, the measured
conductivity was reported as $0.13231$ $(1/\mu\Omega cm)$, thus
according to Fig. \ref{Fig3}, the thickness could be estimated about
$8 nm$ which is in good agreement with values obtained by direct
experimental measurements for conductivity.

Since, for our sample $s_{0}$ is of the order of $10^{-4}$, and the
average power spectrum of each rough boundary is of the order of
$10^{-2}$, the scattering due to the cross-correlation of the two
interfaces is smaller than the scattering due to each of the rough
surfaces. This issue motivates finding the effect of the
cross-correlation term on conductivity by removing this term from
the conductivity equation (Eq. (\ref{1})). In Fig. \ref{Fig3} graphs
of electrical conductivity of Cu thin film with and without the
coherent term is illustrated. It is evident that the cross
correlation term causes a small shift in conductivity. It could be
understood that the cross correlation effects on conductivity is
inversely proportional to the surface thickness. However, this issue
could be shown by plotting the relative difference of the
conductivity against the thickness, see right panel of Fig.
\ref{Fig3}.
It could be deduced form Fig. 3 that for a layer with thickness $d=1
nm$, the consideration of the cross correlation effects would cause
an increase of about $15$ percent in the conductivity. Whereas, for
a layer with $d=10 nm$, this effect would cause an increase of about
$6$ percent.

\begin{figure}[t]
\centering
\includegraphics[width=12cm,height=8cm]{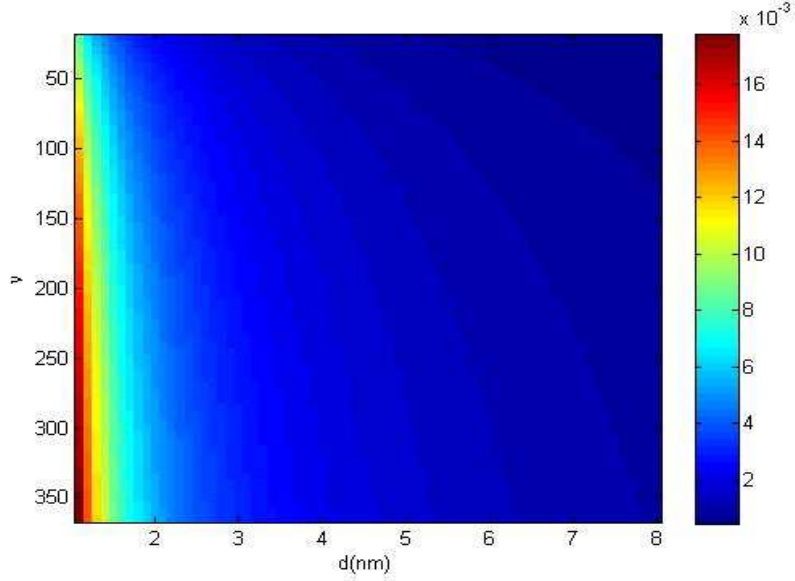}
\caption{Dependence of the capacity resistance $R_{c}$ on the AC
frequency $\nu$ and the layer thickness $d$. The results show how
the capacity increases in small thicknesses and low frequencies.}
\label{Fig4}
\end{figure}

 It is well known that the accumulation of electrons
at sharp points of a surface is more efficient than other places.
Based on this concept the height fluctuations of a rough surface
would prove adequate for accumulation of charges. This leads to the
appearance of local capacities on the rough surface. If a DC current
is applied to the surface due to the fact that the frequency is
infinity, this issue does not affect the electric conductivity. But
if an AC current is applied to the surface, additional resistance is
seen in the system. This resistance is obtained from $R_{c}=(1/C\nu
)$, where $\nu$ is the frequency of the incident wave and $C$ is the
capacitance. The capacitance is obtained by \cite{eom}
\begin{equation}
\label{7} \frac{\langle C\rangle}{\langle
C_{0}\rangle}=1+\frac{2(2\pi)^{5}}{(A_{0})^{2}}\int
(\acute{q})^{3}\langle |h(q)|^{2}\rangle
d\acute{q}+\frac{(2\pi)^{5}}{A_{0}d}\int\coth(\acute{q}d)\acute{q}\langle
|h(q)|^{2}\rangle d\acute{q},
\end{equation}
where $C_{0}$ is the capacitance of two smooth electrode surfaces
$A_{0}$. Note that Eq. (\ref{7}) works for a capacitor with one
rough surface. In our case of study the capacitor consists of two
rough surfaces, this issue obliges further consideration for
obtaining the capacitance. In order to overcome this issue, due to
the assumption that the reciprocal correlation is weak, we may
consider the system as two capacitors in series. Where each
capacitor has one rough surface. This enables (having in hand the
morphology parameters of the surface) measuring the capacitance
effects of the thin Cu layer under an AC current. Equation (\ref{7})
would give the capacitance for the Cu sample by knowing the
roughness parameters. Consequently the capacitance resistance could
be obtained from the capacitance.

Fig. \ref{Fig4} illustrates dependence of the resistivity of the
thin Cu layer on the layer thickness and the frequency of the AC
current applied to the layer. It could readily be noticed that for a
typical layer with specific morphological parameters, by keeping the
frequency constant, the capacitance resistance would be proportional
to the thickness of the layer. This could be explained by the fact
that capacitance is inversely proportional to the thickness of the
layer inside a capacitor. Hence, for thicker layers and lower AC
frequencies, in addition to the Ohmic resistivity, capacitance
resistivity would also come in to play, this is due to the creation
of local capacitors.

\section{Summary}

- An inverse method was proposed to estimate the thin layer
thickness by means of the electric conductivity and the surface
morphology parameters of the layer.

- This method enables to obtain an ensemble average over the various
measurements of the layer thickness by repeating them for different
AC and DC currents. This could reduce the estimation errors of the
layer thickness.

- The reciprocal (cross) correlation of the two rough surfaces of
the thin films increases or decreases the conductivity depending on
its sign (positive sign gives an increase and negative sign gives a
decrease) in comparison to the case where the surfaces where
considered solitary to each other.

- The efficiency of the reciprocal correlation between two rough
surfaces is inversely proportional to the thickness of the layer.
This means that as the thickness of the layer increases the effects
of the reciprocal correlation on the conductivity of the layer is
less efficient.

- An additional resistance to the local resistance of the capacitor
due to the alternate current comes in to play. This resistance which
is named as the capacitance resistance does not exist for the case
of direct current. This is due to the fact that the height
fluctuations create local capacitors on the substrate, and as an
alternate current passes through it an RC circuit is performed.

\textbf{Acknowledgments}: G.R. Jafari would like to thank the
research council of Shahid Beheshti University for financial
support. In addition the authors would like to thank Fatemeh
Iranmanesh for her helpful comments and Sharmin Kharazi for helping
to edit the manuscript.

\end{document}